\documentclass{Interspeech2024}

\interspeechcameraready

\title{Towards Intelligent Speech Assistants in Operating Rooms: A Multimodal Model for Surgical Workflow Analysis}

\name[affiliation={1}]{Kubilay Can}{Demir}
\name[affiliation={1}]{Belén}{Lojo Rodríguez}
\name[affiliation={1}]{Tobias}{Weise}
\name[affiliation={1}]{Andreas}{Maier}
\name[affiliation={1}]{Seung Hee}{Yang}

\address{
  $^1$Friedrich-Alexander-Universität Erlangen-Nürnberg, Germany}
\email{kubilay.c.demir@fau.de}


\begin{document}

\maketitle

\begin{abstract}
To develop intelligent speech assistants and integrate them seamlessly with intra-operative decision-support frameworks, accurate and efficient surgical phase recognition is a prerequisite. In this study, we propose a multimodal framework based on Gated Multimodal Units (GMU) and Multi-Stage Temporal Convolutional Networks (MS-TCN) to recognize surgical phases of port-catheter placement operations. Our method merges speech and image models and uses them separately in different surgical phases. Based on the evaluation of 28 operations, we report a frame-wise accuracy of $92.65 \pm 3.52 \%$ and an F1-score of $92.30 \pm 3.82 \%$. Our results show approximately $10\%$ improvement in both metrics over previous work and validate the effectiveness of integrating multimodal data for the surgical phase recognition task. We further investigate the contribution of individual data channels by comparing mono-modal models with multimodal models. 
\end{abstract}

\section{Introduction}
\label{sec:introduction}

Automatically analyzing surgical workflows at different granularity levels is a fundamental task for developing speech assistants in operating rooms (OR). The necessity to develop smart assistants arises from the fact that surgical operations are becoming more complex and demanding procedures. Thus, the number of medical personnel in the operating room and their responsibilities are increasing constantly~\cite{maktabi2017online}. An intelligent system can take over routine tasks or assist medical personnel during procedures~\cite{cleary2004or2020}. These context-aware support systems could be advantageous in managing lights and surgical tables, pre-setting correct configurations on devices for upcoming steps, providing insights about the patient's medical history, or even preparing operative reports. Furthermore, the administration can monitor the progress of all ORs in a central system and prepare the next patients most effectively~\cite{maier2017surgical, padoy2019machine}. These envisioned capabilities of future ORs are now closer to becoming a reality with the advancements in Natural Language Processing (NLP). 

Although surgical operations are very complex procedures, similar patterns can be observed in the same type of operations~\cite{padoy2008line}. These patterns can be explained by predefined surgical actions. The most commonly used actions are surgical \textit{phases}, \textit{steps}, and \textit{activities}~\cite{lalys2014surgical, mackenzie2001hierarchical}. Surgical phases encompass the primary intervals of intervention, representing the highest level actions performed in the operating room, such as anesthesia, sterilization, or cutting. Steps denote more granular units of action required to achieve the surgical objectives within each phase, and surgical activities are the most fine-grained actions, such as grasping an instrument or aspiration. 

As the main objective of this study, we focus on Surgical Workflow Analysis (SWA) at phase level. The vast majority of studies in recognizing surgical phases have primarily focused on laparoscopic cholecystectomy, removal of the gallbladder, and utilized endoscopic videos as the primary data source~\cite{twinanda2016endonet, jin2017sv, czempiel2020tecno, nwoye2022rendezvous, yi2019hard, shi2022attention, zhang2021surgical}. Cataract surgery and corresponding microscopic videos are the second most common operation type and data modality~\cite{primus2018frame, xia2021against}. Speech and audio data are used in~\cite{guzman2021speech} with Spanish online education videos about laparoscopic cholecystectomy and in~\cite{seibold2022conditional} with the German total hip arthroplasty dataset. Demir et al.~\cite{demir23_interspeech} used three audio channels from physician, assistant, and ambient microphones together with X-ray images in port-catheter placement operations. Huaulmé et al.~\cite{huaulme2021micro} used ambient videos and kinematic data generated during robot-assisted surgeries to recognize surgical phases in anastomosis on artificial blood vessels. 

\begin{figure*}[!htb]
  \centering
  \includegraphics[width=\textwidth]{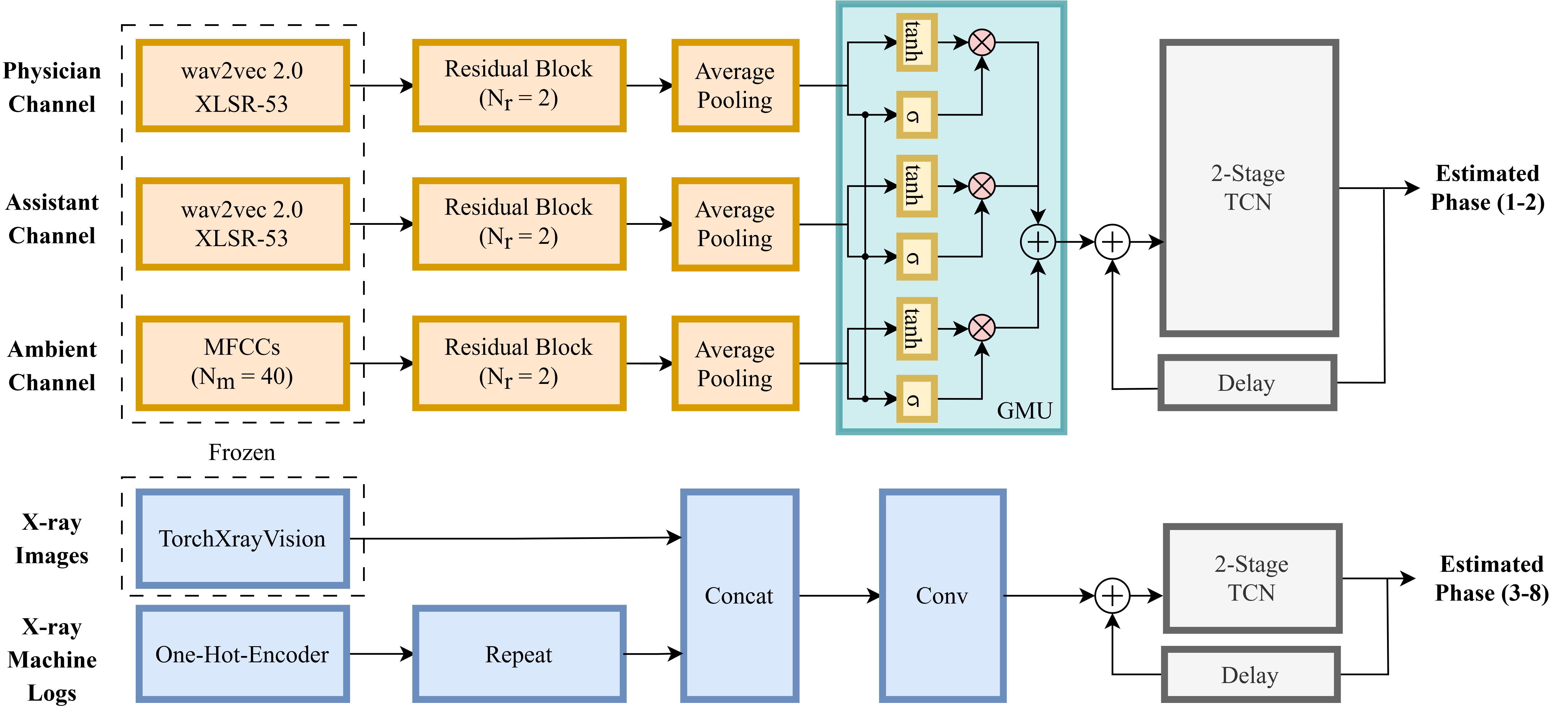}
  \caption{The framework of our proposed model. The top diagram represents the speech model, and the bottom diagram the image model. $N_m$ shows the number of MFCC features, and $N_r$ the number of residual blocks.} 
  \label{fig:model} 
\end{figure*}

Building upon the previous work in~\cite{demir23_interspeech}, we explore an effective way of using multimodality in an SWA task by merging two separate models trained specifically on different phases of port-catheter placement operations. In the feature extraction backbone, our method leverages the wav2vec 2.0 XLSR-53~\cite{conneau2020unsupervised} model in physician and assistant channels, Mel Frequency Cepstral Coefficients (MFCC) in the ambient channel, and TorchXRayVision~\cite{cohen2022torchxrayvision} model in X-ray images. In addition to previous work, we complement X-ray features with an encoded feature vector representing the information from the X-ray machine log files. We employ Gated Multimodal Units (GMU)~\cite{arevalo2017gated} for fusing three audio channels and the Multi-Stage Temporal Convolutional Network (MS-TCN)~\cite{farha2019ms} with an autoregressive connection for modeling temporal relations in both models. The contribution of our work is as follows: 

\begin{itemize}
\item[1)] We propose a multimodal model for recognizing phases of port-catheter placement operations. Our model comprises two distinct speech and image models focusing on different surgical phases, see Figure~\ref{fig:model}.

\item[2)] We extensively evaluate the effectiveness of our approach in the phase recognition task. We further investigate the individual contribution of each data channel within the model.\footnote{ \url{https://github.com/mabel-lr/multimodal-swa}}
\end{itemize}

\section{Method}
\label{sec:method}

\subsection{Feature Extraction}
Generally, the physician and assistant channels are considered as clean, while the ambient channel is considered as noisy and reverberant, as described in Section~\ref{sec:dataset}. In the physician and assistant channels, we employed the wav2vec 2.0 XLSR-53~\cite{conneau2020unsupervised} model for the feature extraction. XLSR-53 is the large multilingual variant of wav2vec 2.0~\cite{baevski2020wav2vec} and is trained with 53 languages including the target language of our experiments, German. Input signals from the entire physician and the assistant channels are windowed using a 10-second window and forwarded to the XLSR-53 model. In experiments, we used the output of the last transformer layer to capture maximum temporal information. In the ambient channel, we extracted 40-dimensional MFCC features. By using MFCC features, we aimed to better represent complementary background noises and non-speech signals present in the channel. 

We utilized the Densenet121~\cite{huang2017densely} model pre-trained on publicly available chest X-ray datasets in the TorchXRayVision library~\cite{cohen2022torchxrayvision} for representing X-ray images. We extracted features from an X-ray image per second and complemented X-ray features with the feature vector created following the X-ray machine log files. The log file provides useful insight into the usage of the X-ray machine with 393 variables. Among those, we utilized three variables to detect: 1) X-ray mode is \textit{fluoroscopy}, 2) X-ray mode is \textit{digital subtraction angiography (DSA)}, 3) X-ray machine is moving. We one-hot-encoded three possible situations of the X-ray machine and repeated each element 64 times after themselves to obtain a 192-dimensional feature vector. 

\subsection{Gated Multimodal Units}
\label{sec:gmu}

The fusion of different modalities is categorized into two approaches: feature-level fusion and decision-level fusion. While feature-level fusion works on combining intermediate representations of modalities, decision-level fusion combines decisions of different models. In this study, we focus on feature-level fusion to integrate three microphone channels using Gated Multimodal Units architecture~\cite{arevalo2017gated}. For an arbitrary input modality feature vector $x_k$, output vector $o_k$ is computed as: 

\begin{align}
\label{eq:gmu_def}
& h_k = tanh(W_{hk} \cdot x_k) \\
& z_k = \sigma (W_{zk} \cdot [x_1,..., x_k]) \\
& o_k = h_k \cdot z_k,
\end{align} 

\noindent where parameters $W_{hk}$ and $W_{zk}$ are learnable weight matrices of the modality $k$, $k \ \epsilon \ [1,K]$, $tanh$ and $\sigma$ are the non-linear tanh and sigmoid functions, and $[\cdot, \cdot]$ is the concatenation operation. The final output of the GMU is obtained by summing output vector $o_k$ of each modality, $h=o_1+...+o_K$. The output vector $h$ is then used as the input vector for the temporal model.

\subsection{Temporal Convolutional Network}
\label{sec:tcn}

To effectively and efficiently analyze temporal relations, we utilize MS-TCN architecture~\cite{farha2019ms}. For a given feature vector $x_{t}$, $t \ \epsilon \ [1,T]$, where $T$ is the duration of an operation in seconds, our goal is to estimate surgical phase $y_t$ at time step $t$. 

In the MS-TCN model, the initial convolutional layer with kernel size $1x1$ maps input feature dimensions to the desired feature dimension. Later, stacked dilated convolutional layers~\cite{oord2016wavenet} and a $1x1$ convolutional layer with a residual connection form a Single-Stage TCN. Finally, stacked Single-Stage TCNs form a Multi-Stage-TCN model. The dilation factor is doubled at each layer of the Single-Stage TCN to increase the receptive field. The dilated residual convolutional layer comprising a dilated convolutional layer, a $1x1$ convolutional layer, and a residual connection are formulated as: 

\begin{align}
\label{eq:tcn}
& \hat{d}_l = ReLU(W_{1,l} * d_{l-1} + b_{1,l}) \\
& d_l =  d_{l-1} + W_{2,l} * \hat{d}_l + b_{2,l},
\end{align}.

\noindent where $d_l$ is the output of the layer $l$, $l \ \epsilon \ [1,L]$ of total $L$ layers, $W_{1,l}$ is the learnable weights of the dilated convolutional layer, $W_{2,l}$ is the learnable weights of the $1x1$ convolutional layer, $b_{1,l}$ and $b_{2,l}$ are bias values, and $*$ is the convolution operation. At the end of each Single-Stage TCN, a $1x1$ convolutional layer with \textit{Softmax} activation function is used to obtain class probabilities. In the MS-TCN, the complete model trained by minimizing the total losses of all Single-Stage TCNs. 

\section{Experiments}
\label{sec:experiments}

\subsection{Dataset}
\label{sec:dataset}

In our experiments, we use the non-public 
\textit{PoCaP Corpus} derived from port-catheter placement operations~\cite{demir2022pocap}. To the best of our knowledge, there is no other multi-modal dataset for any operation type to replicate our experiments similarly. 

Port-catheter placement is a commonly performed operation for patients requiring frequent infusions such as during chemotherapy. The intervention involves placing a port device beneath the chest skin to connect large veins that lead to the heart through a catheter. A single physician and an assistant typically perform the procedure. The audio data from the physician and the assistant are recorded with Sennheiser XSW 2 ME3-E wireless headsets during entire interventions in the dataset. These channels capture the clean speech of medical personnel from a short distance and suppress most of the background noises. The third channel is the ambient channel recorded from the embedded microphone of the GoPro Hero 8 camera installed initially for assisting annotation work. Typically, this channel is reverberant and captures speech from long distances as well as noises coming from medical instruments in the OR. 

The corpus contains recordings of 40 operations annotated with eight surgical phases and one transition phase as in~\cite{demir2022pocap}. 
We used 28 operations in our experiments, with 18/5/5 training, validation, and test set split. The excluded operations suffer from severe data corruption due to hardware or software failure, a change in operating personnel, or a recording mistake. In the data split, we considered the strong variation of surgical phase durations in operations and aimed for three data splits to have the closest possible portion of data for each eight surgical phases. Inspired by the stratified sampling, we used the following steps for splitting sets: (1) Compute the percentage of the phase duration, i.e., the duration of a phase scaled by the total duration of this phase in all operations. (2) Repeat the first step for all phases in all operations. (3) Express each operation using a vector containing eight values, where each value represents the percentage of the phase duration. (4) Compute the entropy of each vector to measure how uniform phases are distributed in an operation. (5) Sort all operations in the dataset according to entropy values in ascending order. (6) Assign the first five operations to the validation set, the next five to the test set, and the remaining operations to the training set. 

\begin{figure}
  \centering
  \includegraphics[width=\linewidth]{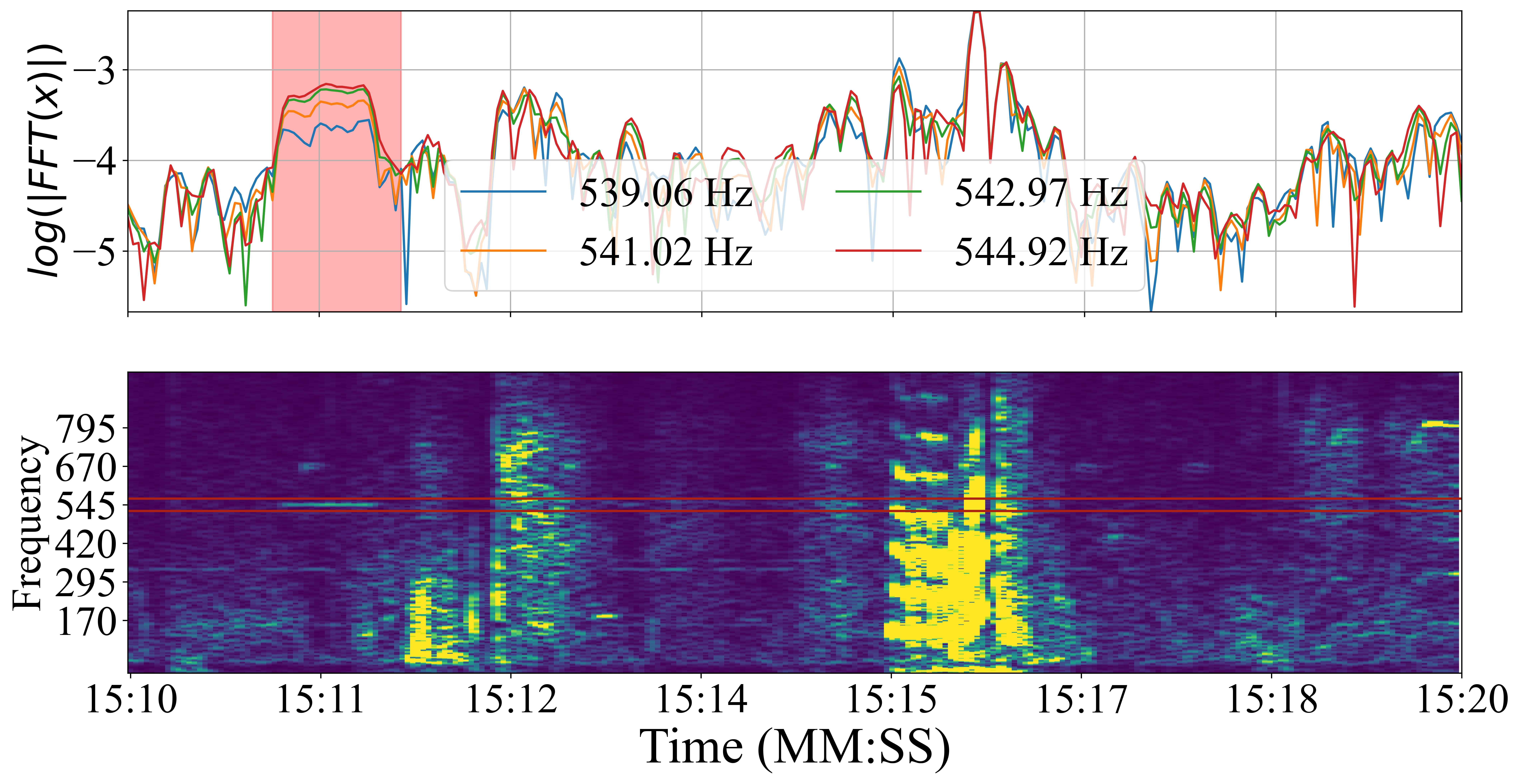}
  \caption{Detecting an audio event of the X-ray machine being active via physician microphone channel. The upper graph shows amplitudes of frequencies, log-scaled for better visualization. The lower graph shows corresponding spectrogram and region of interest with horizontal red lines.}
  \label{fig:align}
\end{figure}

Personal microphone channels and X-ray images are automatically aligned while recorded with the Open Broadcaster Software (OBS)~\cite{Obs2022}. However, the ambient channel is recorded with an external microphone and X-ray machine logs are recorded in a local computer. To align the ambient channel, we used the autocorrelation function (ACF) of the ambient and the physician channels. Then, we used the point of maximum value in ACF to determine the \textit{lag} between. We zero-padded the lagging signal to match with other signals and trimmed the extra tailing parts. If the physician channel is zero-padded, corresponding part in the X-ray channel is filled with black frames, see~\cite{demir2022pocap}. To align X-ray machine logs with the other sources, we used the variable in log files, which indicates that the X-ray machine is producing a warning sound when it is active. To capture this sound in the audio recordings, we used the physician microphone channel and computed amplitudes of frequency responses between $539 Hz$ to $545 Hz$. We detect an audio event if amplitudes of frequencies are consistently growing in this range, see Figure~\ref{fig:align}. Following the audio event detection, we update the time column in the X-ray machine log files to match with the time of the audio events.

\subsection{Model}
As a method of recognizing surgical phases, we propose merging a speech and an image model. The construction of both models is illustrated in Figure~\ref{fig:model}. In the speech model, extracted features are received at each second, refined with residual blocks, and downsampled with a pooling layer to the chosen model feature dimension 256. Then, all input channels are combined with the GMU. Temporal relations modeled with a seven-layer 2-Stage TCN model with autoregressive connection. We verify the effectiveness of the autoregressive connection as in the reference work~\cite{demir23_interspeech} and use it in our model. In the image model, the X-ray image feature vector and X-ray machine logs feature vector are concatenated initially and passed through a convolutional layer. In the temporal model, we use a four-layer 2-Stage TCN model with autoregressive connection. In both models, we use acasual convolutions TCN networks, Label-distribution-aware margin (LDAM) loss~\cite{cao2019learning}, Adam optimizer~\cite{kingma2014adam} with weight decay $1E-6$, learning rate $9E-6$, and batch size $180$ seconds. Although we do not perform estimations in real-time, we assume that the chosen batch size is clinically relevant. Our method is implemented in PyTorch, and models are trained on a single NVIDIA Quadro RTX 4000. 

\begin{figure*}[!tbh]
  \centering
  \includegraphics[width=\textwidth]{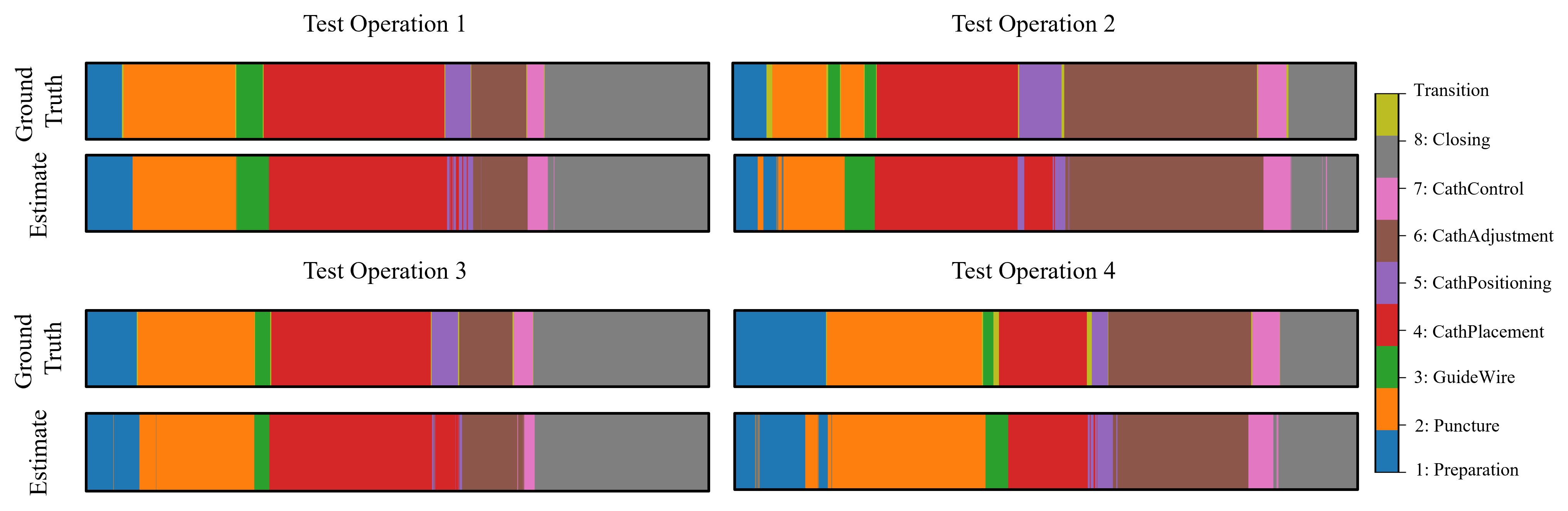}
  \caption{Visualization of estimated surgical phases and ground truth labels of four port-catheter placement operations.}
  \label{fig:ribbon}
\end{figure*}

The choice of our merged model design is motivated by the definitions of surgical phases provided with the dataset~\cite{demir2022pocap}. We observe that the X-ray features are not relevant for recognizing the Preparation and Puncture phases. Meanwhile, noises while covering the patient with the sterile towel and utterances of the physician before the local anesthesia are prominent points for recognizing these phases. In contrast, the Positioning of the Guide Wire and Catheter Control phases are initiated or terminated with movements of the X-ray machine. Fluoroscopy in Positioning of the Guide Wire, Catheter Positioning, and Catheter Adjustment phases have different durations and focus on various regions or instruments. Moreover, the DSA is only used during the Catheter Control phase. With these observations, we design our model such that the speech model estimates the Preparation and Puncture phases, while the image model estimates phases from Positioning of the Guide Wire to Closing. We switch from the speech model to the image model during the Puncture phase if the speech model outputs consistent estimations for a selected duration. The integration of prior knowledge and exploiting the potential of multimodality allows us to focus on smaller but simpler problems to solve the initially intricate problem. 

\subsection{Results}

To evaluate the results of the phase recognition task, we deploy frame-wise accuracy and macro-averaged F1-Score metrics. The accuracy indicates the proportion of frames accurately classified based on the ground truth labels and the F1 score represents the harmonic mean of precision and recall. Macro-averaging assumes equal importance in all phases and provides a more comprehensive understanding of the model's performance due to the imbalanced class distribution. We repeat each experiment three times and report the average values of the mean and standard deviation of each metric. 

\begin{table}[h]
  \caption{Comparison of the phase recognition results between the reference model and our proposed model.}
  \label{tab:reference_comparison}
  \centering
  \begin{tabular}{l c c}
    \toprule
    \textbf{Model} & \textbf{Accuracy(\%)} & \textbf{F1-Score(\%)} \\
    \midrule
    \textit{PoCaPNet~\cite{demir23_interspeech}} & $82.56 \pm 3.21$ & $81.30 \pm 3.89$        \\
    Ours & $92.65 \pm 3.52$ & $92.30 \pm 3.82$ \\
    \bottomrule
  \end{tabular}
\end{table}

In Table~\ref{tab:reference_comparison}, we present the performance of our model and the comparison with the reference work. Comparing our proposed method with PoCaPNet~\cite{demir23_interspeech}, we notice approximately $10\%$ improvement in both metrics. In Figure~\ref{fig:ribbon}, we visualize the predictions for four operations. The results clearly show the capabilities of our method and the consistent phase estimations. In addition to the overall recognition accuracy improvement, another major contribution is related to the Catheter Positioning (purple) phase, which was consistently not recognized previously. Our method improves on this issue significantly and detects a large portion of the phase. However, it suffers from over-segmentation error, which we will address in future studies. Although the durations of all phases are not distributed equally, they have equal clinical relevance, and it is vital to recognize all phases. The Catheter Positioning phase has the shortest average duration among all phases and substantial variations in implementation during operations. Therefore, it is challenging to recognize this phase robustly. 

\begin{table}[h]
  \caption{Phase recognition results of individual channels within speech and image models.}
  \label{tab:channels}
  \centering
  \begin{tabular}{l c c}
    \toprule
    \textbf{Channels} & \textbf{Accuracy(\%)} & \textbf{F1-Score(\%)} \\
    \midrule
    Speech-All& $93.08 \pm 3.49$ & $93.17 \pm 3.42$ \\
    \hspace{1mm} $\cdot$ Only Physician& $88.99 \pm 2.84$ & $88.62 \pm 3.30$ \\
    \hspace{1mm} $\cdot$ Only Assistant& $74.35 \pm 7.19$ & $66.77 \pm 11.60$ \\
    \hspace{1mm} $\cdot$ Only Ambient& $81.39 \pm 6.36$ & $79.73 \pm 7.77$ \\
    \hline
    Image-All& $92.34 \pm 4.60$ & $91.81 \pm 4.84$ \\
    \hspace{1mm} $\cdot$ Only X-ray& $81.14 \pm 9.05$ & $76.38 \pm 9.17$ \\
    \bottomrule
  \end{tabular}
\end{table}

In Table~\ref{tab:channels}, we further investigate the contribution of individual channels in speech and image models. In parallel to our hypothesis in model design to incorporate multiple data channels, we achieved the best results when using all available channels, see \textit{Speech-All} and \textit{Image-All}. When training the speech model on individual channels, we observe the best results when using only the physician channel as input. However, the performance significantly drops when the assistant channel is used. In data recordings, we observe that medical assistants might leave the OR, chat with other personnel, or deal with other tasks between their duties in an operation. Therefore, the performance drop in the assistant channel is consistent with our observation. The ambient channel shows a $9\%$ drop in F1-Score compared to the physician channel. Although an ambient microphone can theoretically capture all speech and audio data in the OR, this data channel is very noisy and reverberant. Thus, the contribution is limited. X-ray images are the main channel in the image model. As we only used four variables in log files and considered them as complementary data sources, we did not test this channel separately. However, we achieved superior results by supporting X-ray images with the X-ray machine log files and reported an approximately $15\%$ increase in F1-Score. 

\section{Conclusion}
In this work, we introduced a multimodal framework for the SWA task using three speech and audio channels, X-ray images, and X-ray machine log files. We propose merging a speech and image model to effectively perform surgical phase recognition in different phases. In the speech model, we used the wav2vec 2.0 XLSR-53 model and MFCCs for feature extraction. In the image model, we used Densenet pre-trained on chest X-ray images and one-hot-encoded feature vectors derived from X-ray machine log files. We fused three audio channels with GMU and modeled temporal relations with MS-TCN architectures. We report approximately $10\%$ improvement compared to the previous work and analyze the contribution of individual data channels in port-catheter placement operations. 

\section{Acknowledgement}
We gratefully acknowledge funding for this study by Friedrich- Alexander-University Erlangen-Nuremberg, Medical Valley e.V. and Siemens Healthineers AG within the d.hip framework.

\bibliographystyle{IEEEtran}
\bibliography{mybib}

\end{document}